\documentclass[pre,twocolumn,aps,showpacs,amsmath,amssymb]{revtex4-1}
\usepackage{color}
\usepackage{bm}
\usepackage{graphicx}
\usepackage{braket}
\usepackage[plainpages=false,pdfpagelabels,colorlinks=true,linkcolor=red,urlcolor=blue,citecolor=blue,pdftitle={Title},pdfauthor={},pdfdisplaydoctitle=true,pdfduplex=DuplexFlipLongEdge]{hyperref}
\def\Tr{\mbox{Tr}\,}

\begin{document}
\title{Quantum adiabatic protocols using emergent local Hamiltonians}
\author{Ranjan Modak} 
\author{Lev Vidmar} 
\author{Marcos Rigol}
\affiliation{Department of Physics, Pennsylvania State University, University Park, Pennsylvania 16802, USA}

\begin{abstract}
We present two applications of emergent local Hamiltonians to speed up quantum adiabatic protocols for isolated noninteracting and weakly interacting fermionic systems in one-dimensional lattices. We demonstrate how to extract maximal work from initial band-insulating states, and how to adiabatically transfer systems from linear and harmonic traps into box traps. Our protocols consist of two stages. The first one involves a free expansion followed by a quench to an emergent local Hamiltonian. In the second stage, the emergent local Hamiltonian is ``turned off'' quasistatically. For the adiabatic transfer from a harmonic trap, we consider both zero- and nonzero-temperature initial states.
\end{abstract}

\maketitle

\section{Introduction} 

The field of far-from-equilibrium dynamics in isolated quantum many-body systems has attracted great interest in recent years, addressing old and opening new fundamental questions in quantum mechanics, statistical physics, and quantum information~\cite{d2016quantum, borgonovi_santos_16, eisert_friesdorf_15, polkovnikov_sengupta_11}. As a result, theoretical concepts such as entanglement generation after quantum quenches~\cite{calabrese_cardy_05, calabrese_cardy_07, eisler07, alba_calabrese_17}, the generalized Gibbs ensemble (GGE) in integrable systems~\cite{rigol2007relaxation, calabrese_essler_11, ilievski_denardis_15, vidmar_rigol_16, essler_fagotti_16, cazalilla_chung_16, caux_16}, and eigenstate thermalization in quantum chaotic systems~\cite{deutsch1991quantum, srednicki1994chaos, rigol2008thermalization, rigol2012alternatives}, have been established and used to gain an understanding of a wide range of nonequilibrium phenomena. 

Extraordinary advances in experiments with ultracold quantum gases are an important driving force in this progress~\cite{greiner02, kinoshita2006quantum, bloch_dalibard_08, cazalilla_11}. They have created unique setups for the exploration of strongly correlated many-body quantum systems in and out of equilibrium. Ultracold quantum gases are usually inhomogeneous because of the presence of confining potentials that are, to a good approximation, harmonic. A quest to prepare homogeneous systems is currently underway to realize and study quantum phases of interest and their transitions~\cite{mukherjee_yan_17, mazurenko_chiu_17}. One of the quantum adiabatic protocols considered in this work is motivated by this quest.

The last two decades have also witnessed much interest in developing a thermodynamic framework for small and nonequilibrium quantum systems. Fluctuation theorems~\cite{jarzynski1997nonequilibrium, crooks1999entropy, kurchan2000quantum, tasaki2000jarzynski, campisi2011colloquium} and information theory~\cite{aaberg2013truly, horodecki2013fundamental} have become useful tools for the exploration of work extraction in the context of quantum thermodynamics. More recently, there have been studies that connect developments in the understanding of the dynamics of isolated quantum systems with those in quantum thermodynamics \cite{perarnau2016work, modak.2017, verstraelen2017unitary, mblengine17}. In Ref.~\cite{modak.2017}, two of us discussed how to extract maximal work by means of quantum quenches and quasistatic processes in isolated noninteracting (described using a GGE) and weakly interacting [described using the grand canonical ensemble (GE)] fermionic systems in one-dimensional (1D) lattices. A quantum adiabatic protocol considered here is motivated by the goal of extracting maximal work and reducing the time required to extract it.

A challenge for the current generation of nonequilibrium studies is to apply existing knowledge of quantum engineering and controlled manipulation to design adiabatic protocols for many-body systems. Here we are interested, in particular, in using nonequilibrium dynamics to speed up such protocols. This topic is not new. It has been discussed, mostly at the single-particle level, within the framework of the so-called shortcuts to adiabaticity~\cite{Torrontegui13}. One of the most common ideas explored in this context is the use of counterdiabatic drivings~\cite{demirplak2003adiabatic, demirplak2005assisted, deffner.2014, campbell.2015, sels_polkovnikov_17}, in which a time-dependent Hamiltonian is used to achieve adiabatic dynamics.

Here we tackle the challenge of using nonequilibrium dynamics to speed up adiabatic transformations by employing the recently introduced concept of emergent eigenstate solutions to quantum dynamics~\cite{vidmar.2017}. Emergent eigenstate solutions, and their associated emergent Gibbs ensembles \cite{vidmar_xu_17}, have been used to explain a dynamical quasicondensation phenomenon~\cite{vidmar.2017, rigol04, vidmar.2015} and effective cooling during expansion dynamics \cite{xu_rigol_17}. In this work, we show that the emergent eigenstate solution also provides a framework to generate shortcuts to adiabaticity. The cornerstone of our approach is the construction of an emergent local Hamiltonian, an explicitly time-dependent operator, of which time-evolving (under a time-independent Hamiltonian) pure states are eigenstates. Consequently, no entropy is generated during the nonequilibrium dynamics in the eigenbasis of the emergent local Hamiltonian. Being local, this Hamiltonian can potentially be engineered in a variety of systems.

\begin{figure*}
\centering
\includegraphics[width=0.99\textwidth]{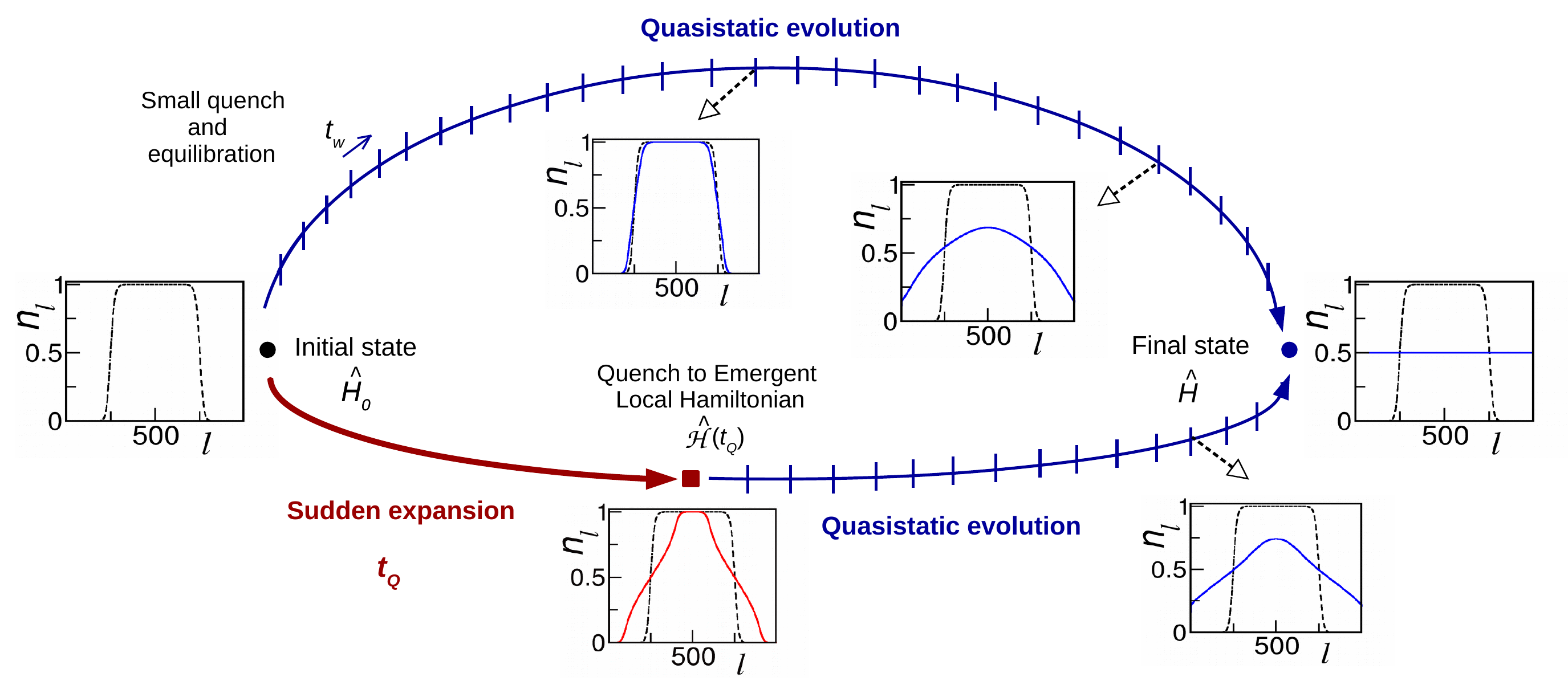}
\caption{Quantum adiabatic protocols discussed in Sec.~\ref{secVb}, in which the initial state is a finite-temperature state in a harmonic trap. We refer to the initial Hamiltonian as $\hat H_0$. The lower path shows the two-stage protocol, which consists of (i) a free expansion for time $t_Q$ followed by a quench to the emergent local Hamiltonian $\hat{\cal H}(t_Q)$, and (ii) a quasistatic process in which $\hat{\cal H}(t_Q)$ is ``turned off'' in $N_s$ small quenches (in the sketch, $N_s=15$), and the system equilibrates after each small quench. We use $t_W$ to denote  the average waiting time between two consecutive small quenches, and we denote the final Hamiltonian as $\hat H$. The upper path (in the sketch, $N_s = 30$) shows the protocol in which the initial trap is turned off quasistatically.  The curves (solid lines) depict the site occupations $n_l =  \langle \hat c_l^\dagger \hat c_l \rangle$ in the GGE at different points in each protocol. Dashed lines show the site occupations in the initial state. Despite the difference in $N_s$ (the total time of both protocols increases approximately linearly in $N_s$), we achieve a similar degree of adiabaticity in both protocols (the energy of the final state, relative to the ideal adiabatic transfer, is 0.85 for the lower path and 0.88 for the upper one).}
\label{fig1}
\end{figure*}

We present two applications of the emergent local Hamiltonian. In the first one, we discuss how to  extract maximal work for initial (filled and empty) band-insulating states. As a second application, we discuss how to adiabatically transfer initial equilibrium states from linear and harmonic traps onto a box trap (a homogeneous lattice with open boundary conditions). In both cases, a faster adiabatic protocol is implemented by allowing the particles to expand freely up to times at which they almost reach the edge(s) of the empty part(s) of the lattice. At that point, the appropriate emergent local Hamiltonian is quenched, so that the expanding state freezes (this occurs because the expanding state is either the ground state or a Gibbs state of the emergent local Hamiltonian). We then ``turn off'' the emergent local Hamiltonian in a quasistatic fashion using a sequence of small quenches, and letting the system equilibrate after each small quench. The equilibration processes are the ones taking the overwhelming majority of  time in our protocols. The key steps are presented in Fig.~\ref{fig1}.

We study the degree of adiabaticity achieved as a function of the time at which the emergent local Hamiltonian is quenched and of the number of small quenches used. The total time of the protocol increases approximately linearly with the number of small quenches. As an extreme case, we compare the results of the two-stage protocol with the straightforward quasistatic turn off of the initial trapping potential. Figure~\ref{fig1} shows the site occupations at different points in our protocols, for an initial finite-temperature state in a harmonic trap (see Sec.~\ref{secVb}). To achieve a similar degree of adiabaticity, a considerably smaller number of small quenches is needed in the two-stage protocol.

The paper is organized as follows. In Sec.~\ref{secII}, we introduce the protocols and statistical ensembles used in the calculations. The first application involving work extraction is presented in Sec.~\ref{secIII}, while Secs.~\ref{secIV} and~\ref{secV} are devoted to the adiabatic transfers from linear and harmonic traps, respectively, to a box trap. We summarize our results in Sec.~\ref{secVI}.

\section{Quantum adiabatic protocols\label{secII}}

We consider initial states that are spatially inhomogeneous in lattices that contain unoccupied sites. Those states are taken to be either ground states or finite-temperature states of a Hamiltonian $\hat H_0$. (We define $\hat H_0$ separately for the applications studied in Secs.~\ref{secIII}--\ref{secV}.) The dynamics of initially inhomogeneous states in 1D lattices has recently attracted much interest both for fermionic models~\cite{hm08, hm09, kajala11, bolech12, vidmar13, mei16, herbrych_feiguin_17} and quantum spin chains (or hard-core bosons)~\cite{antal99, rigol04, rigol_muramatsu_05a, gobert05, eisler09, lancaster10, sabetta_misguich_13, alba14, vidmar.2017, lancaster16, bertini16, eisler_maislinger_16, vidmar_xu_17, xu_rigol_17, deluca_collura_17, ljubotina_znidaric_17, kormos_17, collura_deluca_17, alba_17,stephan_17}.

The quantum adiabatic protocols implemented in our work are split into two stages (see also Fig.~\ref{fig1}). The first stage consists of a sudden expansion and a quench to an emergent local Hamiltonian (see Sec.~\ref{secIIa}), and the second stage is a quasistatic evolution (see Sec.~\ref{secIIb}).

\subsection{Sudden expansion and the emergent local Hamiltonian} \label{secIIa}

During the sudden expansion, the initial state expands under the free (1D) Hamiltonian
\begin{eqnarray}
\hat{H}&=&- J\sum _{l=1}^{L-1}(\hat{c}^{\dag}_l\hat{c}^{}_{l+1}+\text{H.c.}) \, ,
\label{h0}
\end{eqnarray}
where $\hat{c}^{\dag}_{l}$ ($\hat{c}^{}_{l}$) is the fermionic creation (annihilation) operator at site $l$, and $L$ is the number of lattice sites. In what follows, we set the hopping amplitude $J$ to unity ($J$ sets our energy scale).

After an expansion time $t_Q$, we quench $\hat H\rightarrow\hat{\cal H}(t_Q)$, where $\hat{\cal H}(t_Q)$ is the emergent local Hamiltonian~\cite{vidmar.2017}:
\begin{equation} \label{def_Heme}
 \hat{\cal H}(t_Q) = e^{-i t_Q \hat H} \hat H_0 e^{i t_Q \hat H} \,
\end{equation}
where we have set $\hbar=1$. While the definition of the latter operator appears to be simple, the crucial property that we require for $\hat{\cal H}(t_Q)$ is {\it locality}, i.e., $\hat{\cal H}(t_Q)$ must be an extensive sum of operators with support on $O(1)$ lattice sites. It is not immediately obvious that $\hat{\cal H}(t_Q)$ can be a local operator. In fact, even for solvable models, this is generically not the case. The locality of $\hat{\cal H}(t_Q)$ follows from the commutation relations between $\hat H$ and $\hat H_0$ upon expanding Eq.~(\ref{def_Heme}) in a power series of $t_Q$. In general, $\hat{\cal H}(t_Q)$ is a local operator if the nested commutators of $\hat H$ with $\hat H_0$ vanish at some order, or they close the sum~\cite{vidmar.2017}. The families of quantum quenches for which local emergent Hamiltonians $\hat{\cal H}(t_Q)$ have been constructed include the following: (i) $\hat H_0$ is a boost operator for $\hat H$, relevant to quadratic models and anisotropic Heisenberg chains~\cite{vidmar.2017}, and (ii) $\hat H_0$ contains $\hat H$ and a power-law potential that is turned off at the quench, relevant to quadratic models~\cite{vidmar_xu_17}. The latter is the one used in this work.

The existence of an emergent local Hamiltonian results in some remarkable consequences for the quantum dynamics. If the initial state is the (nondegenerate) ground state of $\hat H_0$, then the state at time $t_Q$ is the (nondegenerate) ground state of a local operator $\hat{\cal H}(t_Q)$~\cite{vidmar.2017}. If the initial state is a Gibbs state of $\hat H_0$ with inverse temperature $\beta$, then the state at time $t_Q$ is a Gibbs state of a local operator $\hat{\cal H}(t_Q)$ with inverse temperature $\beta$~\cite{vidmar_xu_17, xu_rigol_17}. As a result of the quench to $\hat{\cal H}(t_Q)$, the expansion stops (the system ``freezes'') because the time-evolving state is a stationary state of $\hat{\cal H}(t_Q)$.

\subsection{Quasistatic evolution} \label{secIIb}

In the second stage, which starts at time $t_Q$, we transform $\hat{\cal H}(t_Q)$ into $\hat{H}$ by means of a quasistatic process (we also refer to this transformation as a ``turn off'' of the emergent local Hamiltonian). The quasistatic process is a unitary time evolution that consists of $N_s$ ``small'' quantum quenches. We use the word ``small'' to emphasize that $N_s$ is typically large and hence the excess energy induced by a single quench is small. In the limit $N_s \to \infty$, the ground state of the emergent local Hamiltonian at time $t_Q$ is transformed into the ground state of $\hat H$ at the end of the protocol.

While the $N_s$-step quasistatic transformation of $\hat{\cal H}(t_Q) \to \hat H$ can be performed in many different ways, here we focus on very simple protocols. Generally, we turn off the relevant parameters linearly. Namely, if a parameter $\eta$ is to be set to zero, this is achieved by means of $\eta_{n_s} = \eta (1-n_s/N_s)$, where $n_s = 1,...,N_s$. The only exception is when turning off a harmonic trap (studied in Sec.~\ref{secV}). For that case we consider two options: turning off the trap amplitude linearly, or turning off the characteristic density (which is proportional to the square root of the trap amplitude)~\cite{rigol_muramatsu_04c} linearly.

The system is allowed to equilibrate after each small quench. We denote the average waiting time between two consecutive small quenches as $t_W$. The total time of the protocol is hence
\begin{equation}
 t_{\rm total} = t_Q + N_s \, t_W \, .
\end{equation}
Typically, both $t_Q$ and $t_W$ are proportional to the system size $L$. Actually, the average equilibration time after the small quenches is generally longer than or about the same as the expansion time (see Sec.~\ref{secIIIa}). Therefore, for large $N_s$, the quasistatic evolution takes the overwhelming majority of time in our protocols.

\subsection{Statistical ensembles}

After each small quench in the quasistatic evolution, observables after relaxation can be described by a proper statistical ensemble (to be defined below). Actually, in the context of work extraction in Sec.~\ref{secIII}, we show that nearly indistinguishable results are obtained when unitarily evolving states after equilibration following each small quench are replaced by the density matrix of the appropriate statistical ensemble. We then only apply the statistical ensemble description of the quasistatic evolution in Secs.~\ref{secIV} and~\ref{secV}.

For the strict noninteracting (integrable) evolution, the appropriate statistical ensemble is the GGE, which takes into account an extensive number of nontrivial conserved quantities that prevent thermalization~\cite{rigol2007relaxation, vidmar_rigol_16, essler_fagotti_16, cazalilla_chung_16, caux_16}. The GGE density matrix, which is obtained maximizing the entropy subject to the constraints associated with the nontrivial conserved quantities, can be written as~\cite{rigol2007relaxation}
\begin{equation} \label{def_GGE}
 \hat{\rho}^\text{GGE}=\frac{1}{Z_\text{GGE}} e^{-\sum_\alpha\lambda_\alpha\hat{I}_\alpha} \, ,
\end{equation}
where, for noninteracting spinless fermions, ${\hat{I}_{\alpha}}$, with $\alpha=1,...,L$, are the occupations of the eigenstates of the single-particle Hamiltonian after the small quench, and $Z_\text{GGE}=\Tr[\exp(-\sum_\alpha\lambda _\alpha \hat{I}_\alpha)]$ is the partition function of the GGE. The Lagrange multipliers $\lambda_\alpha$, which are determined by the condition $\Tr[\hat{\rho}^\text{GGE}\hat{I}_{\alpha}] = I_{\alpha} \equiv \Tr[\hat{\rho} \,\hat{I}_{\alpha}]$, can be written as~\cite{rigol2007relaxation}
\begin{equation} \label{Ikgge}
\lambda_\alpha = \ln\left(\frac{1- I_\alpha}{I_\alpha} \right),
\end{equation}
where $\hat{\rho}$ is the density matrix of the state at the time of the small quench.

The GGE entropy is computed as~\cite{he2012initial}
\begin{equation}
S^\text{GGE}=-\sum_{\alpha=1}^{L}[I_\alpha \ln I_\alpha +(1-I_\alpha) \ln(1-I_\alpha)]\,.
\label{eq:sGGE}
\end{equation}

In the presence of very weak integrability-breaking interactions, large systems are expected to thermalize after the small quench~\cite{rigol_14, bertini_essler_15, brandino_caux_15, rigol2016fundamental, bertini_essler_16, mierzejewski_prosen_15}. This is the case even if the interactions are not strong enough to change the expectation value of macroscopic observables from the thermal ones in the noninteracting limit. In such systems, the density matrix that characterizes the state after equilibration can be taken to be the GE one,
\begin{equation} \label{def_GE}
 \hat\rho^\text{GE}= \exp(-\beta [\hat H' - \mu \hat{N}])/Z,
\end{equation}
where $\hat H'$ is the Hamiltonian after the small quench, $\hat{N}$ is the particle number operator (we deal with systems in which $\hat{H}'$ and $\hat{N}$ commute), $\beta$ and $\mu$ are the inverse temperature and chemical potential, respectively, and $Z=\Tr[\exp(-\beta [\hat H'-\mu \hat{N}])]$ is the partition function. $\beta$ and $\mu$ are computed such that the GE energy and number of particles match those in the system undergoing unitary evolution after the small quench.

For the protocols studied in this work (see also Ref.~\cite{modak.2017}), the results obtained using the GGE and GE descriptions are qualitatively similar. Hence, for the sake of brevity, we focus on the GGE description after small quenches. Only in Sec.~\ref{secV}, in which we study the adiabatic transfer of equilibrium states from harmonic traps to box traps, do we present results both for the GGE and the GE descriptions. This is the protocol that is most relevant to current experiments with ultracold gases.

\section{Work extraction\label{secIII}}

In this section we study work extraction, for which it is essential that we generate an initial state that is nonpassive~\cite{pusz1978passive, lenard1978thermodynamical}. We extract work in the following way: (i) We connect two chains with $L/2$ sites by allowing particles to hop between them (the hopping matrix element between them is taken to be $J=1$). This creates a single chain with $L$ sites. The initial state has the form
\begin{eqnarray}
 |\psi_{I}\rangle=|\psi_{I}\rangle_1\otimes|\psi_{I}\rangle_2 ,
 \label{is1}
\end{eqnarray}
i.e., it is a direct product of pure states in chains 1 ($|\psi_{I}\rangle_1$) and 2 ($|\psi_{I}\rangle_2$). We focus on the case in which 
\begin{eqnarray}
|\psi_{I}\rangle_1=\prod_{l=1}^{L/2}\hat{c}^{\dag}_l|\emptyset\rangle_1 \, ,\text{\quad and \quad} |\psi_{I}\rangle_2=|\emptyset\rangle_2,
\label{is2}
\end{eqnarray}
namely, in chain 1 (2) we have a filled (empty) band insulator. (ii) We carry out (nearly) adiabatic unitary transformations (in a system that now has $L$ sites) as prescribed in the two stages mentioned in Sec.~\ref{secII}. (iii) We disconnect the two subsystems with $L/2$ sites to have two disconnected chains as in the initial state. The number of particles ($N=L/2$) remains constant in the entire system at all times.

The work extracted in the cycle, $W$, is defined as
\begin{equation}\label{eq:work}
W=\Tr\left[(\hat{\rho}^I-\hat{\rho}^F)\,(\hat{H}_1+\hat{H}_2)\right],
\end{equation}
where $\hat{H}_1$ ($\hat{H}_2$) is the Hamiltonian of chain 1 (2), and $\hat{\rho}^I=|\psi_{I}\rangle\langle\psi_{I}|$ ($\hat{\rho}^F$) is the density matrix of the initial (final) state. The Hamiltonians $\hat{H}_1$ and $\hat{H}_2$ are given by Eq.~\eqref{h0}, with sums between $l=1$ and $L/2-1$ for $\hat{H}_1$ and between $l=L/2+1$ and $L-1$ for $\hat{H}_2$. In our definition, $W$ is the difference between the energy of the initial and final states~\cite{alicki2013entanglement, d2016quantum}. As a result, $W$ increases as one lowers the energy of the final state.

In a previous work~\cite{modak.2017}, two of us implemented cyclic protocols involving a sudden quench and a quasistatic process that allowed one to extract maximal work in similar setups (we considered both relaxation to the GGE and the GE after a quench). Here, using the emergent local Hamiltonian, we show that not only can one extract maximal work but also, by changing the free expansion time $t_Q$, one can speed up the protocol by reducing the number of small quenches.

Our initial state [see Eqs.~\eqref{is1} and~\eqref{is2}] is an eigenstate of any Hamiltonian that is a sum of site occupation operators with arbitrary coefficients. In particular, it is the ground state of $\hat{H}_0=(1/L)\sum _{l=1}^{L}l \,\hat{n}_l$, where $\hat{n}_l=\hat{c}^{\dag}_{l}\hat c^{}_{l}$ is the occupation operator for site $l$. The expansion dynamics of this state after the two chains are connected is studied under the Hamiltonian $\hat H$ in Eq.~\eqref{h0}. The time-evolving state is the ground state of the emergent local Hamiltonian
\begin{equation}
\hat{\mathcal H}(t)=-\sum _{l=1}^{L-1}(e^{i\pi/2}\hat{c}^{\dag}_l\hat{c}^{}_{l+1}+\text{H.c.}) + \frac{1}{t}\sum_{l=1}^{L} l \, \hat{n}_l \, ,
\label{h1}
\end{equation}
where we have rescaled the Hamiltonian in Eq.~(\ref{def_Heme}) by $\hat{\mathcal H}(t) \to \hat{\mathcal H}(t)L/t$. The time-evolving state is the ground state of $\hat{\mathcal H}(t)$ as long as the propagating front of particles (holes) does not reach the right (left) boundary of our lattice~\cite{vidmar.2017}. This occurs at a time $t_\text{max}\approx L/4$, because the propagating front has to travel $L/2$ sites and the maximal group velocity in the lattice is $2aJ/\hbar$, which is nothing but 2 in our units (we set the lattice spacing $a$ to unity). We quench to the emergent local Hamiltonian at different times $t_Q<t_\text{max}$ to stop the expansion dynamics.  
 
Next, we transform $\hat{\mathcal H}(t_Q)$ [Eq.~\eqref{h1}] into $\hat H$ [Eq.~\eqref{h0}] by means of a quasistatic process, i.e., we perform $N_s$ small quenches. In Eq.~\eqref{h1}, we linearly turn off the trap amplitude $t_Q^{-1} \to 0$ and at the same time we linearly turn off the phase $\pi/2 \to 0$. Since the initial state of the quasistatic process is the ground state of $\hat{\mathcal H}(t_Q)$, then as $N_s\rightarrow\infty$ the system must be the ground state of $\hat H$ at the end of the process. To complete the cycle, and have two independent chains with $L/2$ sites as in the initial state, we disconnect the two halves of the lattice by setting the hopping matrix element between them to zero. This local quench produces an $O(1)$ change of the energy (which is negligible for large system sizes \cite{modak.2017}). Since our initial state has $\langle\psi_{I}|\hat{H}_1+\hat{H}_2|\psi_{I}\rangle=0$, the maximal work that can be extracted in a cycle is the negative of twice the ground-state energy of each chain with $L/2$ sites and $L/4$ particles.

We consider two types of protocol descriptions. In the first one (see Sec.~\ref{secIIIa}), we calculate the unitary time evolution of the wave function after each small quench. In the second one (see Sec.~\ref{secIIIb}), we take the equilibrated state after each small quench to be described by the GGE density matrix.

\subsection{Exact unitary dynamics} \label{secIIIa}

\begin{figure}[!t]
\centering
\includegraphics[width=0.48\textwidth]{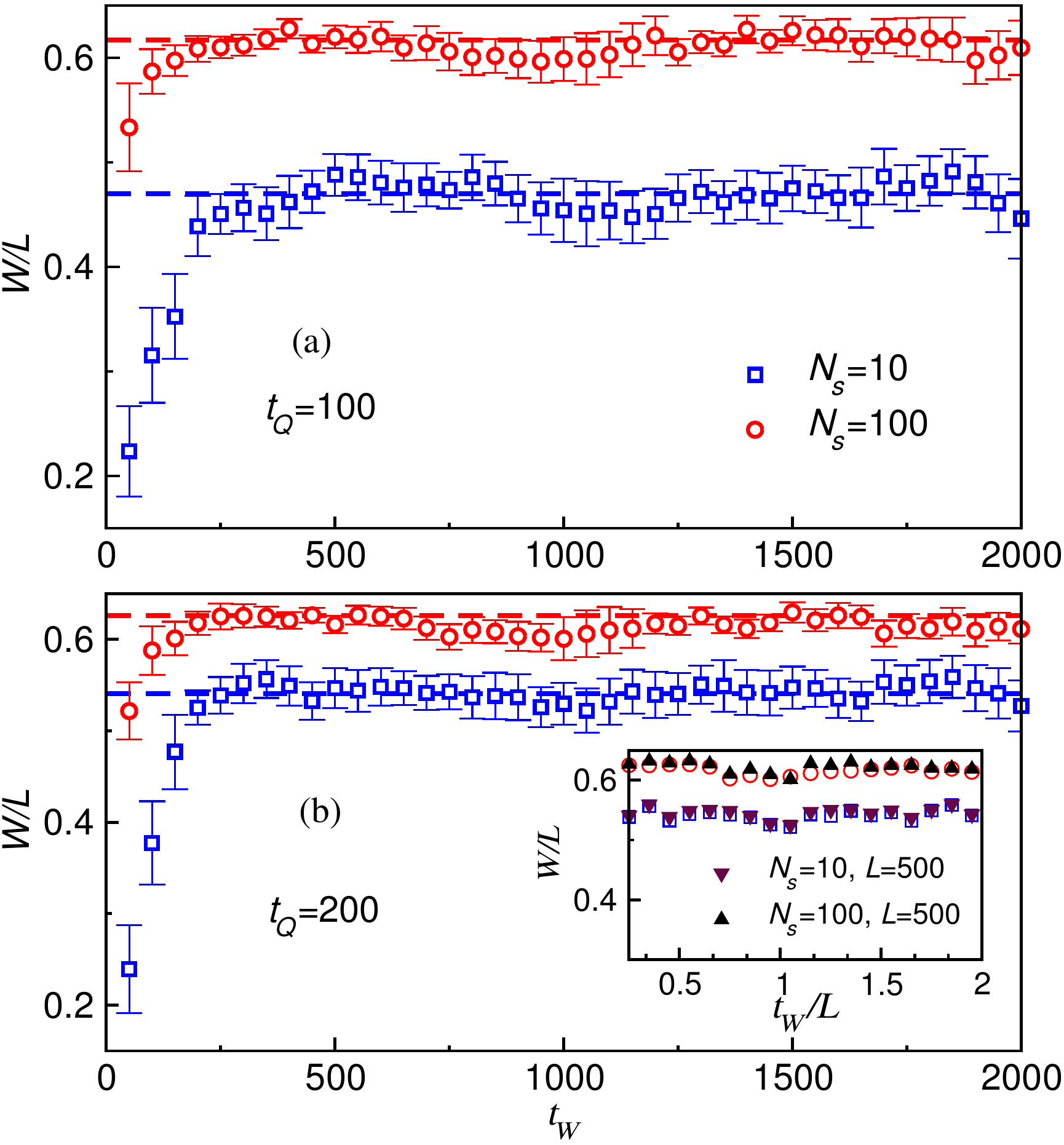}
\caption{Average work extracted per site $W/L$ at the end of the protocol as a function of the average waiting time $t_W$ after each small quench, for $L=1000$ ($N=L/2$). The exact waiting time after each small quench is randomly chosen from the interval $[t_W-t_W/8,t_W+t_W/8]$ with uniform probability. The results are averaged over 500 (100) realizations of the protocol for $N_s=10$ (100), and the error bars denote the standard deviation. Horizontal lines are the GGE predictions. (a) $t_Q=100$, (b) $t_Q = 200$. The inset in (b) shows data collapse for $W/L$ as a function of $t_W/L$ for $L=500$ and $L=1000$ (same results and symbols as in the main panel, $t_Q = 200$).}
\label{rf1}
\end{figure}

\begin{figure}[!t]
\centering
\includegraphics[width=0.48\textwidth]{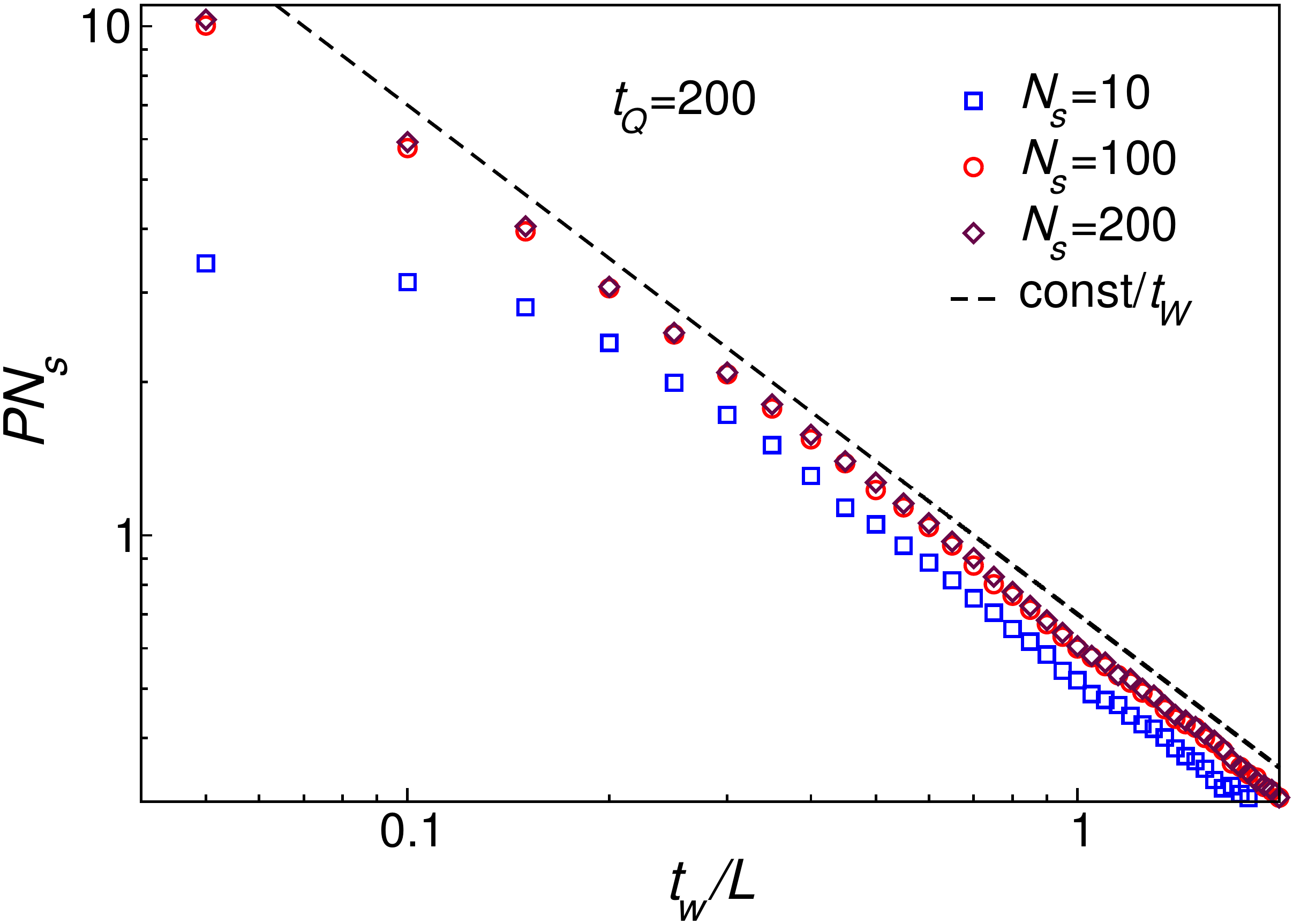}
\caption{Rescaled average power $P \, N_s$, see Eq.~(\ref{def_power}), as a function of the average waiting time $t_W$. Results are shown for $t_Q=200$, $L=1000$ ($N=500$), and different values of $N_s$. We average over 500, 100, and 50 realizations of the protocol for $N_s=10$, 100, and 200, respectively. The standard deviation is not included for clarity.
The dashed line is a function proportional to $t_W^{-1}$.
}
\label{rf2}
\end{figure}

Here we consider the exact unitary dynamics. After each small quench, the system wavefunction evolves for a time that is chosen randomly from an interval $[t_W -t_W/8, t_W+t_W/8]$. This is done to remove coherences that result from the integrability of the system \cite{d2016quantum}. In Fig.~\ref{rf1}, we plot the average extracted work per site $W/L$ as a function of the average waiting time $t_W$. Remarkably, for sufficiently long average waiting times ($t_W\gtrsim t_Q$), $W/L$ fluctuates about the GGE prediction (dashed lines in Fig.~\ref{rf1}, to be discussed in Sec.~\ref{secIIIb}). The inset of Fig.~\ref{rf1}(b) shows data collapse for $W/L$ as a function $t_W/L$ for different values of $L$ (we choose $L=500$ and 1000). Hence, both the work and the average waiting time to achieve a particular work per lattice site, scale with the system size $L$. This is expected for quenches in the inhomogeneous systems studied here.

Next, we discuss the power $P = W/t_{\rm total} = W/(t_Q+N_st_W)$ that can be extracted from our protocol. In the limit $N_s\to\infty$ (ideal adiabatic evolution), the power vanishes as expected, while for finite $N_s$ and $t_W$ it is non-zero. For large $N_s$, one can express $P$ as
\begin{equation} \label{def_power}
 P = \frac{1}{N_s} \frac{1}{t_W} \frac{W}{\left(1 + N_s^{-1} \frac{t_Q}{t_W}\right)} \approx N_s^{-1} (t_W/L)^{-1} \, \frac{W}{L} \, ,
\end{equation}
where we assumed $t_Q/t_W \ll N_s$. In Fig.~\ref{rf2}, we plot $P \, N_s$ versus $t_W/L$ for $t_Q = 200$ and three different values of $N_s$ ($N_s=10$, 100, and 200). The results show that $P N_s \propto (t_W/L)^{-1}$ at large $t_W$ and $N_s \gtrsim 100$. This is reasonable considering that, in this parameter regime, $W/L$ depends only weakly on $t_W$ and $N_s$ [see Figs.~\ref{rf1} and~\ref{fig2}(a), respectively]. For the power $P$ to be large, one should then select the smallest $t_W$ required for the system to equilibrate to the GGE, and vary $N_s$ to reach the desired compromise between maximizing work and power.

\subsection{Generalized Gibbs ensemble} \label{secIIIb}

\begin{figure}[!t]
\centering
\includegraphics[width=0.48\textwidth]{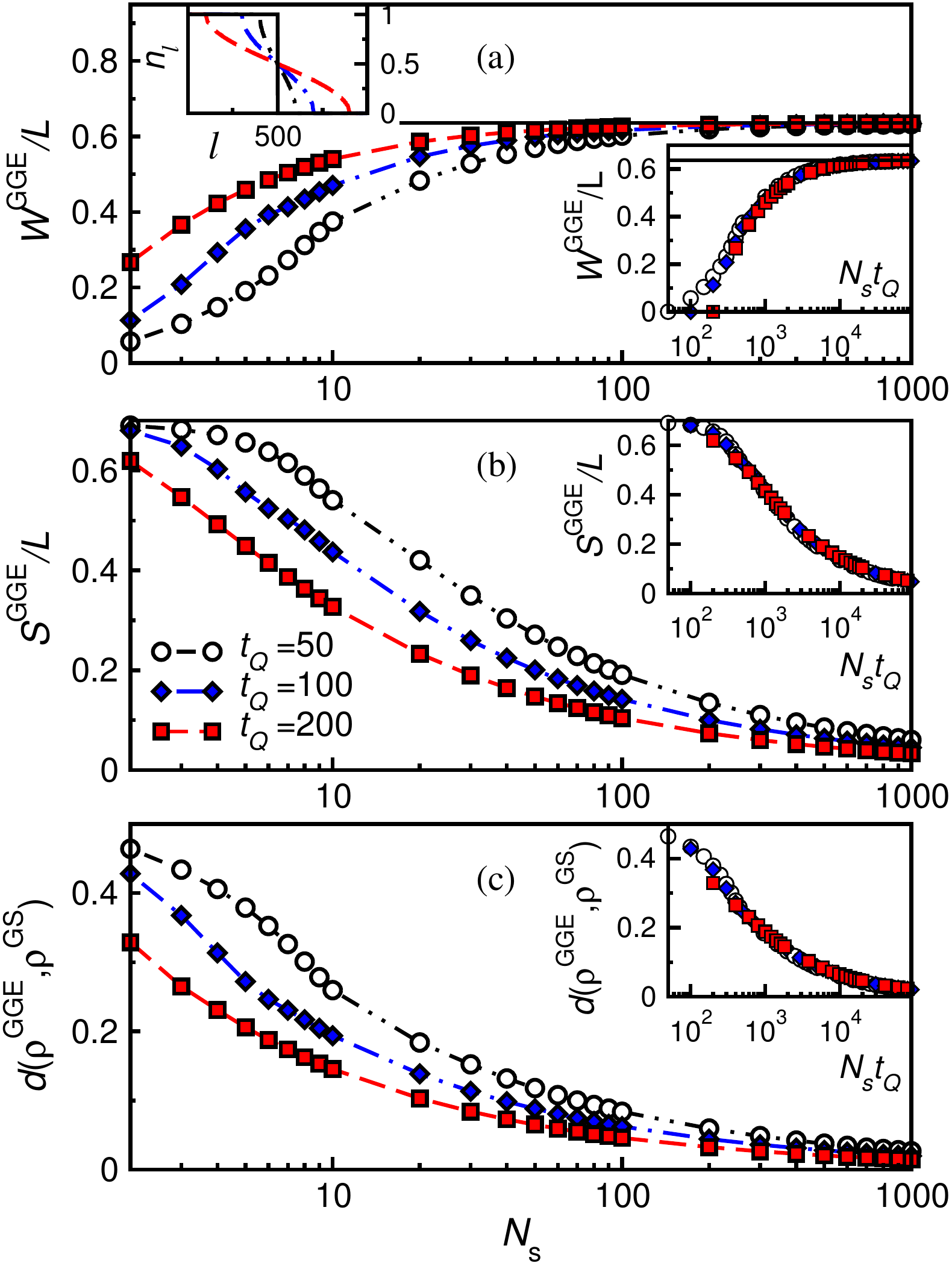}
\caption{(a) Work extracted per site $W^{\rm GGE}/L$ vs the total number of small quenches $N_s$ for three times $t_Q$ at which the emergent local Hamiltonian $\hat{\cal H}(t)$, Eq.~(\ref{h1}), is quenched. The horizontal solid line shows the maximal work that can be extracted (see the text). Left inset: site occupations at the times $t_Q$. The solid line corresponds to the site occupations in the initial state. Right inset: data collapse for $W^{\rm GGE}/L$ as a function of $N_st_Q$. (b) The GGE entropy per site $S^{\text{GGE}}/L$ at the end of the cycle vs $N_s$. Inset: data collapse for $S^{\text{GGE}}/L$ as a function of $N_st_Q$. (c) Normalized trace distance, Eq.~\eqref{td0}, between the GGE one-body correlation matrix at the end of the cycle and that of the ground state of $\hat{H}_1+\hat{H}_2$ vs $N_s$. Inset: data collapse for the trace distance as a function of $N_st_Q$. The results are for $L=1000$ ($N=500$), and for times $t_Q=50$, 100, and 200.}
\label{fig2}
\end{figure}
 
Here we consider the description in which the equilibrated state after each small quench in the quasistatic evolution is replaced by the appropriate GGE density matrix (as justified in Sec.~\ref{secIIIa}).

In Fig.~\ref{fig2}(a), we plot the work extracted per site $W^{\rm GGE}/L$ versus $N_s$ for three times $t_Q$ at which the local emergent Hamiltonian is quenched (for $L=1000$). The left inset in Fig.~\ref{fig2}(a) shows the site occupations $n_l = \langle \hat n_l \rangle$ at those times ($t_Q=50$, $100$ and $200$). Since $t_Q<L/4=250$, the site occupations at the right (left) boundary remain zero (one). The results make it apparent that, as $t_Q$ increases, the work extracted for any given number of small quenches $N_s$ increases. In other words, as $t_Q$ increases, one approaches the maximal work that can be extracted (see the horizontal solid line) more rapidly with increasing $N_s$. Similarly, as $t_Q$ increases, Fig.~\ref{fig2}(b) shows that the GGE entropy per site $S^{\text{GGE}}/L$ at the end of a cycle [see Eq.~(\ref{eq:sGGE})] decreases more rapidly with increasing $N_s$. Note that, consistent with the fact that the entropy vanishes in the ground state, $S^{\text{GGE}}/L$ can be seen to vanish as $N_s\rightarrow\infty$.

To characterize the approach towards the ground state of $\hat{H}_1+\hat{H}_2$, we also calculate the normalized trace distance between the GGE one-body correlation matrix at the end of a cycle and that of the ground state (GS) of $\hat{H}_1+\hat{H}_2$. The normalized trace distance per lattice site is defined as
\begin{eqnarray}
d(\rho^{\text{GGE}},\rho^\text{GS})=\frac{1}{L}\Tr\big[\sqrt{[\rho_{}^{\text{GGE}} -\rho^\text{GS}]^2}\big] \, ,
\label{td0}
\end{eqnarray}
where the matrix elements of the one-body correlation matrix after a cycle are $\rho_{jl}^{\text{GGE}}=\langle \hat c^{\dag}_{j} \hat c^{}_l\rangle_{\text{GGE}}$ and the corresponding ones of the ground state $|\psi_\text{GS}\rangle$ of $\hat{H}_1+\hat{H}_2$ are $\rho_{jl}^{\text{GS}}=\langle\psi_\text{GS}| \hat c^{\dag}_{j} \hat c^{}_l|\psi_\text{GS}\rangle$. Figure~\ref{fig2}(c) shows that, as expected, $d(\rho^{\text{GGE}},\rho^\text{GS})$ vanishes as $N_s\rightarrow\infty$. For any given $N_s$, the normalized trace distance decreases with increasing $t_Q$. 
 
The right insets in Figs.~\ref{fig2}(a)--\ref{fig2}(c) show that, remarkably, if one rescales $N_s$ in the $x$-axes of the main panels multiplying by $t_Q$, all the data for different values of $t_Q$ collapse onto single curves. This means that, in order to achieve a given degree of adiabaticity in our cycles, the number of small quenches required is inversely proportional to the expansion time, for $t_Q<t_\text{max}$. Since most of the time in each cycle is spent in the quasistatic process, the free expansion and quench to the emergent local Hamiltonian at $t_Q\lesssim t_\text{max}$ results in a significant speed up of the cycle. The rescaling obtained can be intuitively understood from the fact that, in the emergent local Hamiltonian in Eq.~(\ref{h1}), the strength of the linear trap is proportional to $1/t$. This means that the longer the free expansion time (for $t_Q<t_\text{max}$), the weaker is the trap that one needs to turn off in a quasistatic fashion.

\section{Adiabatic transfer from a linear trap to a box trap\label{secIV}}

\begin{figure}
\centering
\includegraphics[width=0.48\textwidth]{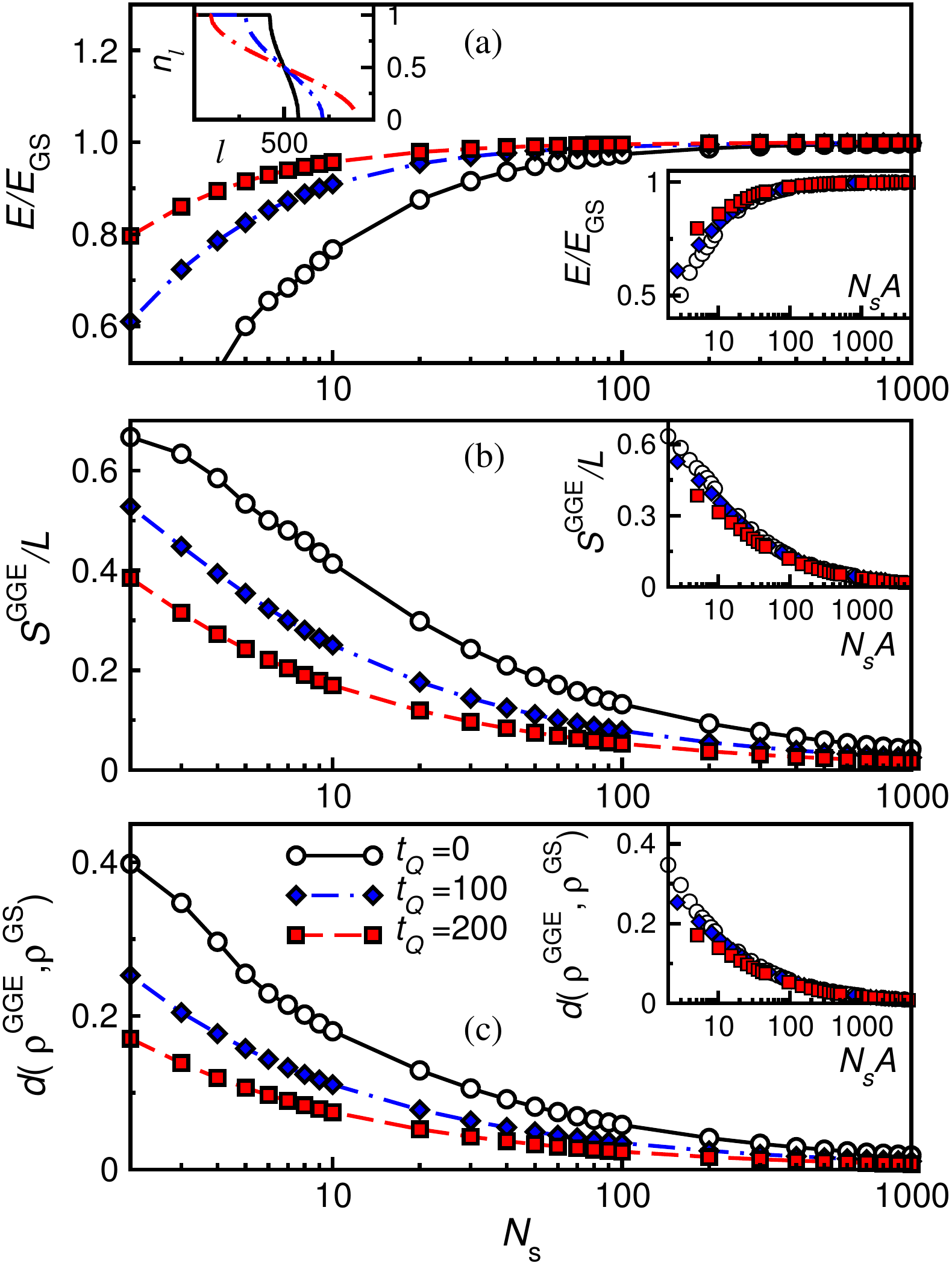}
\caption{Adiabatic transfer of the ground state of a linear trap to a box trap. (a) Ratio between the energy $E$ at the end of the transfer and the ground-state energy $E_\text{GS}$ of $\hat H$ vs the number of small quenches $N_s$ for three times $t_Q$ at which the emergent local Hamiltonian $\hat{\mathcal H}^{(1)}(t_Q)$, Eq.~(\ref{h3}), is quenched. Left inset: site occupations at the times $t_Q$. The solid line corresponds to the site occupations in the initial state. Right inset: data collapse for $E/E_{\text GS}$ as a function of $N_s A(t_Q)$. (b) The GGE entropy per site $S^{\text{GGE}}/L$ at the end of the transfer vs $N_s$. Inset: data collapse for $S^{\text{GGE}}/L$ as a function of $N_s A(t_Q)$. (c) Normalized trace distance $d(\rho^{\rm GGE},\rho^{\rm GS})$ between the GGE one-body correlation matrix at the end of the transfer and that of the ground state of $\hat H$ vs $N_s$. Inset: data collapse for the trace distance as a function of $N_s A(t_Q)$. The results are for $L=1000$ ($N=500$), $\gamma=25$, and for times $t_Q=0$, 100, and 200.}
\label{fig3}
\end{figure}

In this section, we study the adiabatic transfer of particles initially confined in a linear trap into a box trap. We focus on an initial state with $N=L/2$ particles, which is the ground state of the Hamiltonian
\begin{equation}
\hat{H}_{0}^{(1)}=-\sum _{l=1}^{L-1}(\hat{c}^{\dag}_l\hat{c}^{}_{l+1}+\text{H.c.})+\frac{\gamma}{L} \sum _{l=1}^{L}l \, \hat{n}_l \, ,
\label{h2}
\end{equation}
where $\gamma$ is the strength of the confinement. While the initial state is a product state in the limit $\gamma \to \infty$ [the one in Eqs.~(\ref{is1}) and (\ref{is2})], here we are interested in states generated by finite $\gamma > \gamma^* = 4$, so that the site occupations at the right (left) edge of our chain are zero (one) \cite{vidmar.2017}. The left inset of Fig.~\ref{fig3}(a) displays the site occupations $n_l$ in an initial state with $\gamma=25$ ($L=1000$), which we use in the remainder of this section.

The emergent local Hamiltonian for this set up was constructed in Ref.~\cite{vidmar.2017}. Here, we renormalize $\hat{\mathcal H}^{(1)}(t) \to \hat{\mathcal H}^{(1)}(t)/{A}(t)$, where $A(t)=\sqrt{1+(\gamma t/L)^2}$, and we omit an unimportant offset, to obtain
\begin{equation}
\hat{\mathcal H}^{(1)}(t) = -\sum_{l=1}^{L-1} (e^{i\phi(t)} \hat{c}^{\dag}_l\hat{c}^{}_{l+1} + \text{H.c.})+\frac{\gamma}{LA(t)} \sum _{l=1}^{L}l \, \hat{n}_l \, ,
\label{h3}
\end{equation}
where $\phi(t)=\arctan(\gamma t/L)$. The time-evolving state is the ground state of $\hat{\mathcal H}^{(1)}(t)$ as long as the propagating front of particles (holes) does not reach the right (left) lattice boundary. For $\gamma> \gamma^*$, the time at which that occurs in our setup is $t_{\rm max} \approx (L/4)\sqrt{1-(\gamma^{*}/\gamma)^{2}}$~\cite{vidmar.2017}.

After suddenly turning off the linear confining potential [setting $\gamma\rightarrow0$ in Eq.~\eqref{h2}], we follow the two-stage protocol described in Sec.~\ref{secII}. In the first stage, the fermions expand freely under the Hamiltonian $\hat H$ [see Eq.~\eqref{h0}], until a time $t_Q < t_{\rm max}$ at which we suddenly quench $\hat H \to \hat{\mathcal H}^{(1)}(t_Q)$, freezing the expanding cloud. In the second stage, we apply $N_s$ small quenches to set the parameters $\phi(t_Q)$ and $\gamma / [L\mathcal A(t_Q)]$ of the emergent local Hamiltonian $\hat{\mathcal H}^{(1)}(t_Q)$ to zero (in each small quench, those parameters are reduced by $\phi(t_Q)/N_s$ and $\gamma / [L\mathcal A(t_Q)N_s]$, respectively), such that the final Hamiltonian is $\hat H$. After each small quench, the system is assumed to equilibrate to the GGE density matrix. In contrast to the case for which work extraction was studied in Sec.~\ref{secIII}, the purely quasistatic protocol ($t_Q = 0$) is well defined here. The left inset in Fig.~\ref{fig3}(a) shows the site occupations $n_l$ for the three times $t_Q=0$, $100$, and $200$ considered in what follows.

Figure~\ref{fig3}(a) depicts the ratio between the energy $E=\Tr[\hat{\rho}^{F}\hat{H}]$ at the end of the protocol, and the ground state energy $E_\text{GS}$ of $\hat H$. We plot results for three times $t_Q$ as a function of the total number of small quenches $N_s$. As expected, when $N_s \rightarrow \infty$, $E/E_{\rm GS} \to 1$. Figure~\ref{fig3}(b) shows how the GGE entropy per site $S^{\text{GGE}}/L$ [calculated using Eq.~(\ref{eq:sGGE})] at the end of the protocol changes with $N_s$. Increasing $N_s$ reduces the entropy, and our results are consistent with a vanishing value for $N_s \rightarrow \infty$. Results for the normalized trace distance between the one-body correlation matrix of the GGE at the end of the protocol $\rho^{\rm GGE}$ and the one-body correlation matrix $\rho^{\rm GS}$ of the ground state of $\hat H$, $d(\rho^{\rm GGE},\rho^{\rm GS})$ [see Eq.~(\ref{td0})], are shown in Fig.~\ref{fig3}(c). They make it apparent that as $N_s$ increases, $\rho^{\rm GGE}$ approaches $\rho^{\rm GS}$. The results in Figs.~\ref{fig3}(a)-\ref{fig3}(c) reveal that, the larger the value of $t_Q < t_{\rm max}$ is, the smaller is the number of small quenches needed to achieve a desired degree of adiabaticity in the two-stage protocol.
 
The right insets in Figs.~\ref{fig3}(a)--\ref{fig3}(c) show data collapse for the energy, the entropy per site, and the normalized trace distance when rescaling the $x$ axes to $N_s A(t_Q)$. This makes it apparent that, when the protocol includes the initial free expansion, one requires $A(t_Q)$ fewer small quenches in the quasistatic stage to reach the ground state of the box trap with a desired accuracy. The scaling observed can be intuitively understood from the structure of the emergent local Hamiltonian $\hat{\mathcal H}^{(1)}(t)$ in Eq.~(\ref{h3}). In the latter, the strength of the linear confinement $\gamma /[LA(t)]$ weakens as $t$ increases so, as in Sec.~\ref{secIII}, one ends up needing to turn off a weaker trap the longer one waits to stop the free expansion (as long as $t_Q < t_{\rm max}$). Note that, when $\gamma t_Q/L\gg1$, the scaling with $t_Q$ obtained here matches that in Sec.~\ref{secIII}.

\section{Adiabatic transfer from a harmonic trap to a box trap} \label{secV}

In this section, we study the adiabatic transfer of the ground state and a finite-temperature state of a harmonically trapped system into a box trap. The initial Hamiltonian is
\begin{equation}
\hat{H}_{0}^{(2)}=-\sum _{l=1}^{L-1}(\hat{c}^{\dag}_l\hat{c}^{}_{l+1}+\text{H.c.}) + \frac{1}{R^2} \sum _{l=1}^{L}\tilde{l}^{\,2} \, \hat{n}_l \, ,
\label{h4}
\end{equation}
where $\tilde{l}=l-(L+1)/2$ (the center of the trap is in the middle of two sites). The characteristic density, which needs to be kept constant in order to define the finite-density thermodynamic limit in the presence of a harmonic trap~\cite{rigol_muramatsu_04c}, can be written as $\tilde \rho = N/R$. In the ground state of $\hat H_0^{(2)}$, the site occupations in the center of the trap exhibit a band-insulating plateau ($n_l = 1$) when $\tilde\rho \gtrsim 2.6$~\cite{rigol_muramatsu_04c}. In what follows, we set the parameters $\tilde\rho = 10$, $L=1000$, and $N = L/2$. The inset in Fig.~\ref{fig4} (solid line) shows $n_l$ for these parameters.

For the initial finite-temperature state, the density matrix is chosen to be the GE density matrix $\hat \rho^{\rm GE}$ obtained by substituting $\hat H'\to\hat{H}_{0}^{(2)}$, $\beta\to\beta ^{I}$, and $\mu\to\mu^{I}$ in Eq.~(\ref{def_GE}). We take the initial inverse temperature to be $\beta^{I}=0.5$. The chemical potential $\mu^I$ is selected so that $N = L/2$. The entropy of the initial state $S^I$ is calculated using Eq.~\eqref{eq:sGGE} by replacing $I_{\alpha}$ with the occupation of single-particle states in the GE, $I_{\alpha}^{I}=(\exp[\beta^{I}(\epsilon_{\alpha}^{(2)}-\mu^{I})]+1)^{-1}$, where $\epsilon_{\alpha}^{(2)}$ are the single-particle energy eigenvalues of $\hat{H}_0^{(2)}$ [see Eq.~\eqref{h4}].

It was recently shown in Ref.~\cite{vidmar_xu_17} that an emergent local Hamiltonian description can be used to characterize time-evolving states that result from the expansion of initial ground states and finite-temperature states of $\hat{H}_0^{(2)}$. The relevant emergent local Hamiltonian, omitting an unimportant offset, is
\begin{eqnarray}
\hat{\mathcal {H}}^{(2)}(t) & = & \frac{1}{R^2} \sum _{l=1}^{L}\tilde{l}^{\,2} \, \hat{n}_l
- \left( \frac{t}{R} \right)^{\!2} \; \sum _{l=1}^{L-2}(\hat{c}^{\dag}_l\hat{c}^{}_{l+2}+\text{H.c.}) \nonumber \\
& &- \sum _{l=1}^{L-1}A^{(2)}(t,l)(e^{i\phi^{(2)}(t,l)}\hat{c}^{\dag}_l\hat{c}^{}_{l+1}+\text{H.c.}) \, ,
\label{h5}
\end{eqnarray}
where $A^{(2)}(t,l)=\sqrt{1+[(2t/R^2)(l+1/2)]^2}$ and $\phi^{(2)}(t,l)=\arctan\big[2t(l+1/2)/R^2\big]$. For the initial finite-temperature state, the time-evolving density matrix is that of the Gibbs state of $\hat{\mathcal {H}}^{(2)}(t)$ at the inverse temperature $\beta ^{I}$, which we call an emergent Gibbs ensemble~\cite{vidmar_xu_17}. Note that since the temperature of the emergent Gibbs ensemble is identical to that of the initial state, and the emergent and the initial Hamiltonians are related through Eq.~\eqref{def_Heme}, no entropy is generated during the dynamics [$S(t) = S^I$].

The two-stage protocol that we use is similar to the one considered for the initial linear trap studied in Sec.~\ref{secIV}. In the first stage, the particles undergo a free expansion [under the Hamiltonian $\hat H$ in Eq.~(\ref{h0})] into the empty part of the lattice. At time $t_Q$, we suddenly quench $\hat H \to \hat{\cal H}^{(2)}(t_Q)$ to freeze the expanding cloud. We consider $t_Q < t_{\rm max}$, where $t_{\rm max}$ is the time at which the occupation at the boundaries of the lattice depart from zero. (The emergent Hamiltonian description is valid up to that time.) In the second stage, a quasistatic process, we perform $N_s$ small quenches to change $\hat{\cal H}^{(2)}(t_Q) \to \hat H$. Each small quench is followed by an equilibration to the GGE density matrix. In Sec.~\ref{secVb}, we contrast this protocol to the one in which after each small quench the system equilibrates to the GE.

In the quasistatic stage of the protocol, we modify the parameters of the Hamiltonian $(t_Q/R)^2 \to 0$, $\phi^{(2)}(t_Q,l) \to 0$, and $A^{(2)}(t_Q,l) \to 1$ linearly. We consider two ways of turning off the harmonic confinement: (P1) turn off $1/R^2$ linearly, i.e., as $(1-n_s/N_s)/R^2$ where $n_s = 1,...,N_s$, and (P2) turn off the characteristic density $\tilde\rho$ linearly, i.e., turn off the confinement amplitude quadratically as $[(1-n_s/N_s)/R]^2$.

\subsection{Initial ground state} \label{secVa}

\begin{figure}
\centering
\includegraphics[width=0.48\textwidth]{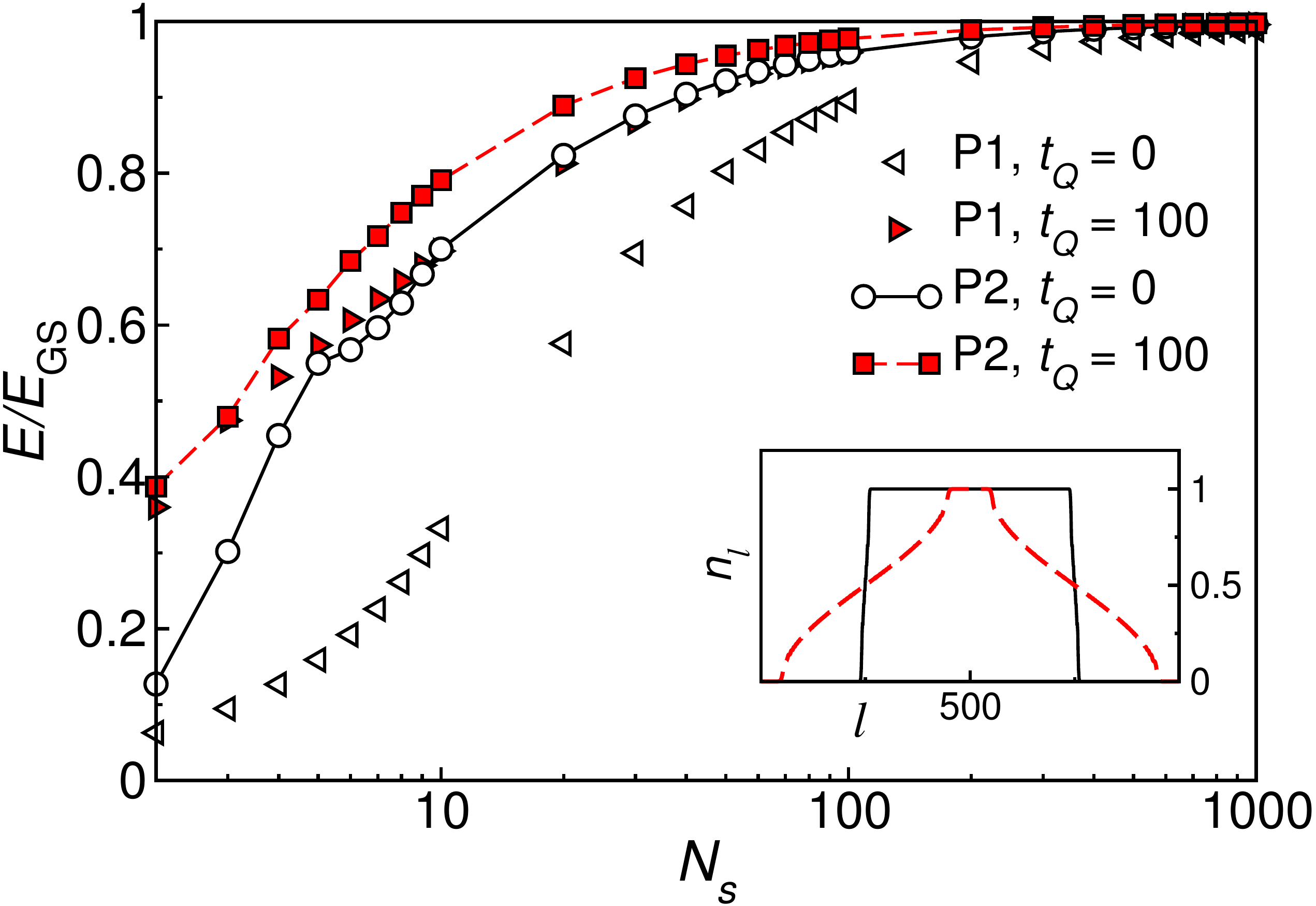}
\caption{Adiabatic transfer of the ground state of a harmonic trap to a box trap.
We compare two ways of turning off the confining potential in the emergent local Hamiltonian $\hat{\cal H}^{(2)}(t_Q)$ [the first term in Eq.~(\ref{h5})] during the quasistatic process. The confining potential is turned off linearly (P1) or quadratically (P2). The main panel shows the ratio between the energy $E$ at the end of the transfer and the ground-state energy $E_{\text{GS}}$ of $\hat H$ vs the number of small quenches $N_s$, for two times $t_Q$ at which $\hat{\cal H}^{(2)}(t_Q)$ is quenched. (Inset) Site occupations at the times $t_Q$. The results are for $L=1000$ ($N=500$), $\tilde{\rho}=10$, and for times $t_Q=0$ and 100.}
\label{fig4}
\end{figure}

First, we consider an initial state that is the ground state of $\hat H_0^{(2)}$ in Eq.~(\ref{h4}), and compare the two different protocols (P1 and P2) to turn off the harmonic confinement (mentioned above). Figure~\ref{fig4} shows the ratio between the energy $E = {\rm Tr}[\hat\rho^F \hat H]$ at the end of each protocol and the ground-state energy $E_\text{GS}$ of $\hat H$ as a function of $N_s$. The results, for free expansion times $t_Q = 0$ and 100, show that the linear turn off of the trap results in a slower approach to the ground state energy as $N_s$ increases when compared to the case of the quadratic turn off (the linear turn off of the characteristic density). 

In the following, we only consider the second (P2) protocol. The linear turn off of the characteristic density (i.e., the quadratic turn off of the harmonic trap) is analogous to the protocol studied for the initial linear trap in Sec.~\ref{secIV}, as the characteristic density in such a potential depends linearly on the strength of the trap. 

The reduction in the number of small quenches $N_s$ required to achieve the same degree of adiabaticity using the sudden expansion is not as good for the initial harmonic trap when compared to the linear one, for the quadratic turn off of the trap (see Fig.~\ref{fig4}). This is likely related to the fact that the emergent local Hamiltonian in Eq.~\eqref{h5} is more complicated than that in Eq.~\eqref{h3}. Still, the reduction with increasing $t_Q$ is apparent in Fig.~\ref{fig4} (note the logarithmic scale of the $x$-axis). Observables such as the entropy and the normalized trace distance at the end of the processes exhibit a behavior (not shown) that is qualitatively similar to that of the energy in Fig.~\ref{fig4}, and to the one discussed in the next section for the initial finite-temperature state.
 
\subsection{Initial finite-temperature state} \label{secVb}

\begin{figure}
\centering
\includegraphics[width=0.48\textwidth]{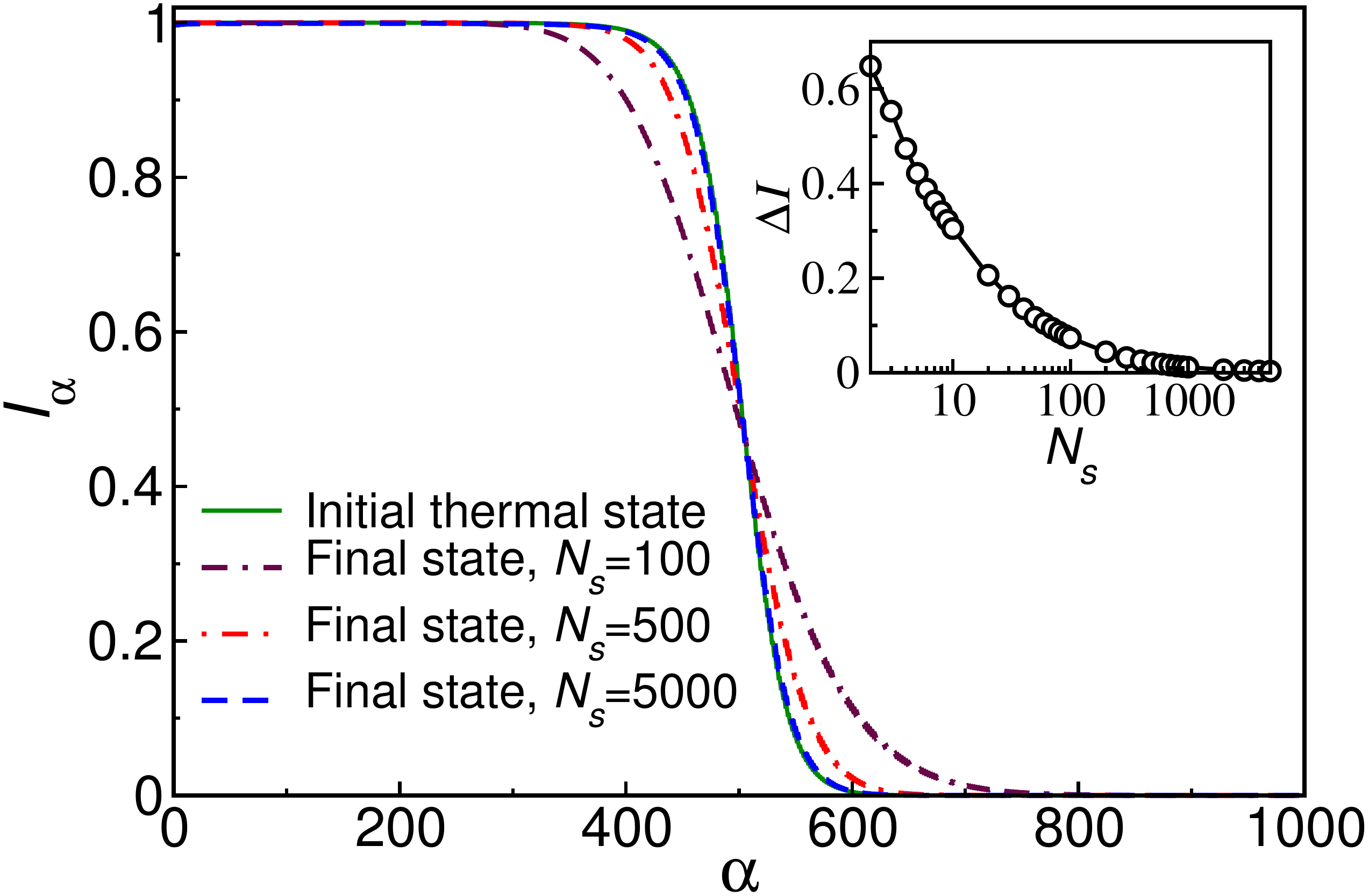}
\caption{Occupations $I_{\alpha}$ of the single-particle energy eigenstates in the initial finite-temperature state ($\beta^{I}=0.5$), and at the end of the two-stage protocol for $t_Q=80$ and different values of $N_s=100$, 500 and 5000. (Inset) Relative difference between the occupation of the single-particle energy eigenstates in the initial and final states (see text) vs $N_s$.}
\label{fig5}
\end{figure} 

Of closer relevance to current experiments with ultracold quantum gases in optical lattices, here we consider an initial finite-temperature state of $\hat H_0^{(2)}$ [see Eq.~(\ref{h4})], and we perform a two-stage adiabatic transfer of this state to a box trap. In the second stage of our protocol (the quasistatic process), after the small quenches, we consider equilibration both to the GGE (as done in all previous cases) and to the GE.

In the limit $N_s\to\infty$ (ideal adiabatic transfer), our protocols do not increase the entropy. Hence, the final entropy of the ideal adiabatic transfer is $S_{\text{adb}}=S^{I}$. When the system equilibrates to the GE, the density matrix at the end of the two-stage protocol, $\hat \rho_{\rm adb}^{\rm GE}$, is uniquely determined by the entropy $S^I$ and the number of particles. When the system equilibrates to the GGE, we find that at the end of the ideal adiabatic transfer, the occupation of the {\em final} single-particle energy eigenstates is the same as the occupation of the {\em initial} single-particle energy eigenstates. This is shown in Fig.~\ref{fig5}, where we plot the occupation of the initial single-particle energy eigenstates $I_{\alpha}^I$, as well as the occupation of the final single-particle energy eigenstates for different numbers $N_s$ of small quenches. The final distribution can be seen to approach the initial one upon increasing $N_s$. The inset in Fig.~\ref{fig5} makes that observation more quantitative. There we plot the relative difference $\Delta I = \sum_\alpha |I_\alpha - I_\alpha^I|/N$ between the occupations of the single-particle energy eigenstates in the initial and final states versus $N_s$. $\Delta I$ can be seen to vanish with increasing $N_s$ (this ensures that, for $N_s\rightarrow\infty$, no entropy is produced within the GGE description). The results shown are for $t_Q=80$, but an identical trend (not shown) was observed for all other times $t_Q$ considered. Hence, in the limit $N_s\to\infty$, the single-particle energy eigenstate occupations are the ones of the initial state, and they uniquely determine the one-body correlation matrix of the ideal adiabatic transfer $\rho_{\text{adb}}^{\rm GGE}$.

\begin{figure}
\centering
\includegraphics[width=0.48\textwidth]{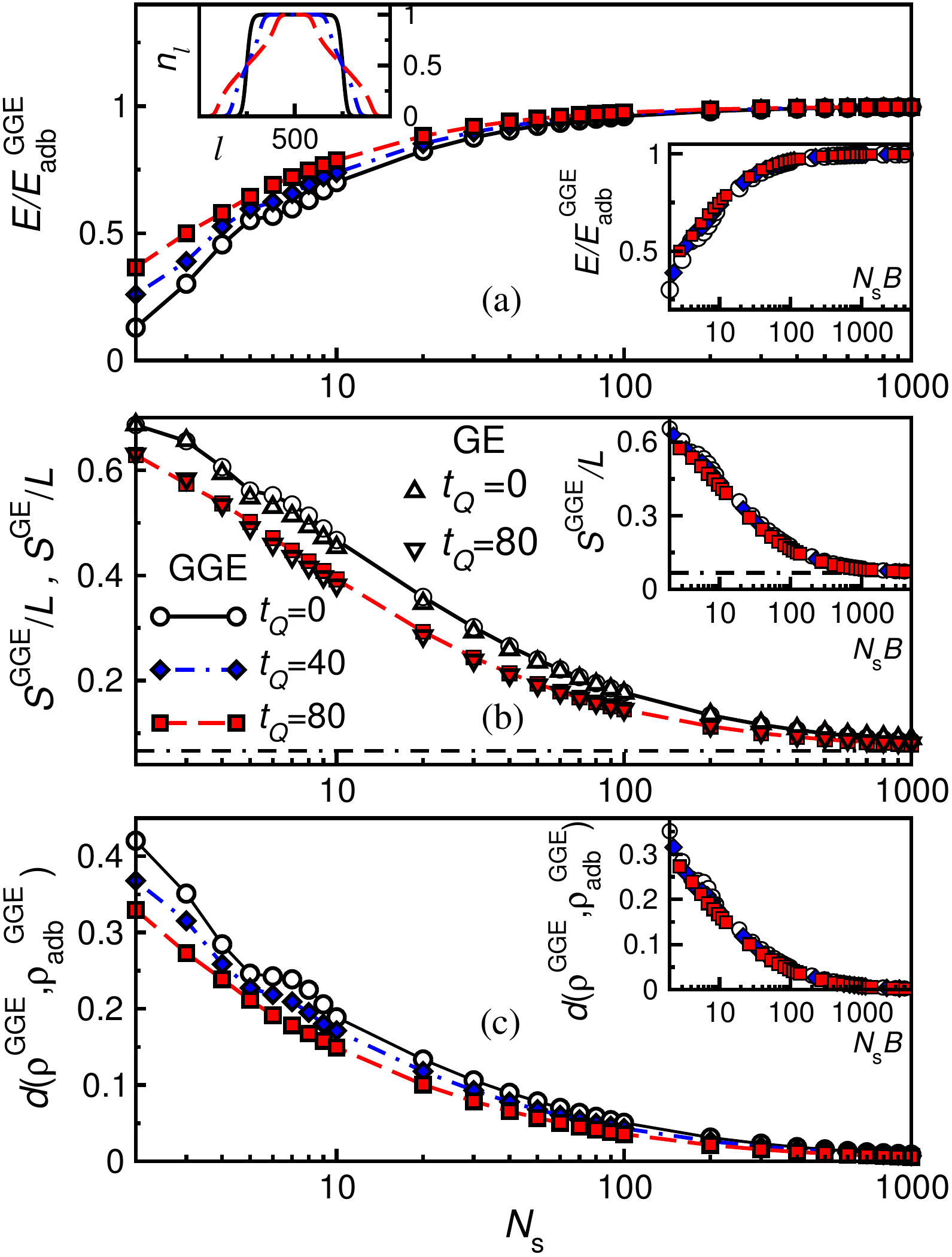}
\caption{Adiabatic transfer of a finite-temperature state from a harmonic trap to a box trap.
(a) Ratio between the energy $E$ at the end of the transfer and the ideal adiabatic transfer energy $E_{\text{adb}}^{\rm GGE}$ vs the number of small quenches $N_s$. Results are shown for three times $t_Q$ at which the emergent local Hamiltonian $\hat{\cal H}^{(2)}(t_Q)$, see Eq.~(\ref{h5}), is quenched. (Left inset) Site occupations at the times $t_Q$. The solid line corresponds to the site occupations in the initial state. (Right inset) Data collapse for $E/E_{\text{adb}}^{\rm GGE}$ as a function of $N_sB(t_Q)$. (b) The GGE entropy per site $S^{\text{GGE}}/L$ at the end of the transfer vs $N_s$. Open triangles correspond to the GE entropy per site $S^{\text{GE}}/L$. The dashed-dotted line corresponds to the entropy $S^{I}/L$ of the initial thermal state. (Inset) Data collapse for $S^{\text{GGE}}/L$ as a function of $N_sB(t_Q)$. (c) Normalized trace distance $d(\rho^{\rm GGE},\rho_{\rm adb}^{\rm GGE})$ between the GGE one-body correlation matrix at the end of the transfer and the one of the ideal adiabatic transfer vs $N_s$. (Inset) Data collapse for the trace distance as a function of $N_sB(t_Q)$. The results are for $L=1000$ ($N=500$), $\tilde{\rho}=10$, $\beta^{I}=0.5$, and for times $t_Q=0$, 40, 80.}
\label{fig6}
\end{figure} 

The left inset of Fig.~\ref{fig6}(a) shows the site occupations $n_l$ for different expansion times $t_Q=0$, $40$ and $80$ (the site occupations at $t_Q = 0$ and 80 are the ones shown in Fig.~\ref{fig1}). The main panels of Fig.~\ref{fig6} show three observables at the end of the transfer, for two or three free expansion times $t_Q$, as a function of $N_s$. In Fig.~\ref{fig6}(a), we plot the ratio between the energy $E$ of the final state and the ideal adiabatic transfer energy $E_{\text{adb}}^{\rm GGE}=\sum_{\alpha}\epsilon_{\alpha} I_{\alpha}^{I}$, where $\epsilon_{\alpha}$ are the single-particle energies of $\hat{H}$. In Fig.~\ref{fig6}(b), we show the GGE entropy per site $S^{\text{GGE}}/L$, calculated using Eq.~(\ref{eq:sGGE}). In Fig.~\ref{fig6}(c), we show the normalized trace distance $d(\rho^{\rm GGE},\rho_{\rm adb}^{\rm GGE})$ between the final GGE one-body correlation matrix $\rho^{\rm GGE}$ and the one-body correlation matrix for the ideal adiabatic transfer $\rho_\text{adb}^{\rm GGE}$, which is obtained replacing $\rho^\text{GS}$ by $\rho_{\text{adb}}^{\rm GGE}$ in Eq.~(\ref{td0}). As a general trend, one can see that: (i) $E \to E_{\rm adb}^{\rm GGE}$, $S^{\rm GGE} \to S^I$, and $d(\rho^{\rm GGE},\rho_{\rm adb}^{\rm GGE}) \to 0$ with increasing $N_s$, and (ii) increasing $t_Q$ decreases the number $N_s$ of small quenches required to achieve a desired degree of adiabaticity during the quasistatic process.

The main panel of Fig.~\ref{fig6}(b) also compares $S^{\text{GGE}}/L$ to $S^{\text{GE}}/L$ at the end of the transfer, when after each small quench the system is assumed to equilibrate to the GE. Even though the final states are different, the results are very close to each other already for relatively small values of $N_s$. This shows that the outcome of the protocol under investigation does not depend significantly  on the choice of the statistical ensemble used to describe the system after equilibration during the quasistatic protocol. This is similar to the results in Ref.~\cite{modak.2017}.

The right insets in Figs.~\ref{fig6}(a)-\ref{fig6}(c) show data collapse for the observables plotted in the main panel as a function of $N_sB(t_Q)$, where $B(t_Q)= [1+(t_Q/R)^2]^{1/4}$. The scaling is different from the one in the linear trap, Sec.~\ref{secIV}. However, the scaling coefficient is, in both cases, an increasing function of the expansion time $t_Q$, so a speed up is achieved whenever the particles are allowed to freely expand into the empty part of the lattice before starting the quasistatic process. 

The protocol discussed in this section could be used in experiments with ultracold gases in optical lattices. In such systems, it might be possible to engineer the emergent local Hamiltonian in Eq.~\eqref{h5}, which contains a harmonic trap, next-nearest-neighbor hoppings with a time-dependent hopping amplitude, and nearest-neighbor hoppings with a time-dependent hopping amplitude and a complex phase. 

\section{Summary} \label{secVI}

We have used emergent local Hamiltonians as a tool to speed up adiabatic protocols for many-body fermionic states in one-dimensional lattices. We focused on two applications of the emergent local Hamiltonians. In the first one, we showed how to extract maximal work from initial band-insulating states. In the second one, we studied the adiabatic transfer of initial equilibrium states from linear and harmonic traps to a box trap. In all the protocols considered, a desired degree of adiabaticity can be achieved using a shorter quasistatic process if one first allows particles to expand freely in the unoccupied part of the lattice and carries out a quench to the emergent local Hamiltonian.  One may wonder why is this so. While we provide no formal proof, one can see that, by using the emergent local Hamiltonian, we manage to ``freeze'' the expanding cloud at times at which the site occupations are nonzero on almost the entire lattice. Hence, after the free expansion the system is much closer to the homogeneous equilibrated state than in the initial state, and this is achieved without producing any entropy. 

Our results demonstrate that the emergent eigenstate solution to quantum dynamics~\cite{vidmar.2017}, and the associated emergent Gibbs ensemble~\cite{vidmar_xu_17}, constitute a promising direction to achieve shortcuts to adiabaticity. It would be interesting to explore its potential further to design quantum heat engines and quantum batteries. We note that the band-insulating states in Sec.~\ref{secIII} can be thought of as being quantum batteries.

\begin{acknowledgements}
This work was supported by the Army Research Office Grant No.~W911NF1410540. The computations were carried out at the Institute for CyberScience at Penn State. We are grateful to Anatoli Polkovnikov for stimulating discussions.
\end{acknowledgements}

\bibliographystyle{biblev1}
\bibliography{reference}

\end{document}